\documentclass[11pt]{article}
\usepackage{fancyhdr}
\def\eqalign#1{\null\,\vcenter{\openup\jot
  \ialign{\strut\hfil$\displaystyle{##}$&$\displaystyle{{}##}$\hfil
      \crcr#1\crcr}}\,}

\usepackage{color}
\usepackage{hyperref}
\usepackage{amsmath}

\usepackage{amsfonts}
\usepackage{graphicx}
\usepackage{epsfig}
\usepackage{caption}

\def\Eq#1{\label{#1}}

\def\iniz{\setcounter{equation}{0}{%
\rhead{\thepage}\lhead{{{{\small\bf\thesection:}
\small \ \SEC\ \  \tiny\today}}}}}

         \let\d=\delta  \let\e=\varepsilon
\let\z=\zeta      \let\th=\vartheta \let\k=\kappa  \let\l=\lambda
    \let\n=\nu     \let\x=\xi        \let\p=\pi     
\let\s=\sigma \let\t=\tau    \let\f=\varphi    
          
\let\G=\Gamma \let\D=\Delta       
\let\P=\Pi

\def\V#1{{\bf#1}}
\def\*{\vskip 3mm}\def\0{\noindent}
\def\be{\begin{equation}}
\def\ee{\end{equation}}
\def\bea{\begin{eqnarray}}
\def\eea{\end{eqnarray}}

\def\NN{{\cal N}}

\def\ie{{\it i.e.}\ }

\def\lis#1{\overline{#1}}
\def\defi{{\buildrel def\over=}}

\def\media#1{{\Blangle\,#1\,\Brangle}}

\def\otto{\,{\kern-1.truept\leftarrow\kern-5.truept\to\kern-1.truept}\,}

\def\tende#1{\,\vtop{\ialign{##\crcr\rightarrowfill\crcr
 \noalign{\kern-1pt\nointerlineskip} \hskip3.pt${\scriptstyle
 #1}$\hskip3.pt\crcr}}\,}
\def\ie{{\it i.e.\ }}

\newdimen\xshift \newdimen\xwidth \newdimen\yshift \newdimen\ywidth%
\def\ins#1#2#3{\vbox to0pt{\kern-#2pt\hbox{\kern#1pt #3}\vss}\nointerlineskip}

\def\eqfig#1#2#3#4#5{
\par\xwidth=#1pt \xshift=\hsize \advance\xshift
by-\xwidth \divide\xshift by 2
\yshift=#2pt \divide\yshift by 2%
{\hglue\xshift \vbox to #2pt{\vfil
#3 \includegraphics{#4.eps}
}\hfill\raise\yshift\hbox{#5}}}

\def\Brangle {{\mbox{\boldmath$ \rangle$}}}
\def\Blangle {{\mbox{\boldmath$ \langle$}}}

  \definecolor{snow}{rgb}{1.000000,0.980392,0.980392}
  \definecolor{ghost white}{rgb}{0.972549,0.972549,1.000000}
  \definecolor{GhostWhite}{rgb}{0.972549,0.972549,1.000000}
  \definecolor{white smoke}{rgb}{0.960784,0.960784,0.960784}
  \definecolor{WhiteSmoke}{rgb}{0.960784,0.960784,0.960784}
  \definecolor{gainsboro}{rgb}{0.862745,0.862745,0.862745}
  \definecolor{floral white}{rgb}{1.000000,0.980392,0.941176}
  \definecolor{FloralWhite}{rgb}{1.000000,0.980392,0.941176}
  \definecolor{old lace}{rgb}{0.992157,0.960784,0.901961}
  \definecolor{OldLace}{rgb}{0.992157,0.960784,0.901961}
  \definecolor{linen}{rgb}{0.980392,0.941176,0.901961}
  \definecolor{antique white}{rgb}{0.980392,0.921569,0.843137}
  \definecolor{AntiqueWhite}{rgb}{0.980392,0.921569,0.843137}
  \definecolor{papaya whip}{rgb}{1.000000,0.937255,0.835294}
  \definecolor{PapayaWhip}{rgb}{1.000000,0.937255,0.835294}
  \definecolor{blanched almond}{rgb}{1.000000,0.921569,0.803922}
  \definecolor{BlanchedAlmond}{rgb}{1.000000,0.921569,0.803922}
  \definecolor{bisque}{rgb}{1.000000,0.894118,0.768627}
  \definecolor{peach puff}{rgb}{1.000000,0.854902,0.725490}
  \definecolor{PeachPuff}{rgb}{1.000000,0.854902,0.725490}
  \definecolor{navajo white}{rgb}{1.000000,0.870588,0.678431}
  \definecolor{NavajoWhite}{rgb}{1.000000,0.870588,0.678431}
  \definecolor{moccasin}{rgb}{1.000000,0.894118,0.709804}
  \definecolor{cornsilk}{rgb}{1.000000,0.972549,0.862745}
  \definecolor{ivory}{rgb}{1.000000,1.000000,0.941176}
  \definecolor{lemon chiffon}{rgb}{1.000000,0.980392,0.803922}
  \definecolor{LemonChiffon}{rgb}{1.000000,0.980392,0.803922}
  \definecolor{seashell}{rgb}{1.000000,0.960784,0.933333}
  \definecolor{honeydew}{rgb}{0.941176,1.000000,0.941176}
  \definecolor{mint cream}{rgb}{0.960784,1.000000,0.980392}
  \definecolor{MintCream}{rgb}{0.960784,1.000000,0.980392}
  \definecolor{azure}{rgb}{0.941176,1.000000,1.000000}
  \definecolor{alice blue}{rgb}{0.941176,0.972549,1.000000}
  \definecolor{AliceBlue}{rgb}{0.941176,0.972549,1.000000}
  \definecolor{lavender}{rgb}{0.901961,0.901961,0.980392}
  \definecolor{lavender blush}{rgb}{1.000000,0.941176,0.960784}
  \definecolor{LavenderBlush}{rgb}{1.000000,0.941176,0.960784}
  \definecolor{misty rose}{rgb}{1.000000,0.894118,0.882353}
  \definecolor{MistyRose}{rgb}{1.000000,0.894118,0.882353}
  \definecolor{white}{rgb}{1.000000,1.000000,1.000000}
  \definecolor{black}{rgb}{0.000000,0.000000,0.000000}
  \definecolor{dark slate gray}{rgb}{0.184314,0.309804,0.309804}
  \definecolor{DarkSlateGray}{rgb}{0.184314,0.309804,0.309804}
  \definecolor{dark slate grey}{rgb}{0.184314,0.309804,0.309804}
  \definecolor{DarkSlateGrey}{rgb}{0.184314,0.309804,0.309804}
  \definecolor{dim gray}{rgb}{0.411765,0.411765,0.411765}
  \definecolor{DimGray}{rgb}{0.411765,0.411765,0.411765}
  \definecolor{dim grey}{rgb}{0.411765,0.411765,0.411765}
  \definecolor{DimGrey}{rgb}{0.411765,0.411765,0.411765}
  \definecolor{slate gray}{rgb}{0.439216,0.501961,0.564706}
  \definecolor{SlateGray}{rgb}{0.439216,0.501961,0.564706}
  \definecolor{slate grey}{rgb}{0.439216,0.501961,0.564706}
  \definecolor{SlateGrey}{rgb}{0.439216,0.501961,0.564706}
  \definecolor{light slate gray}{rgb}{0.466667,0.533333,0.600000}
  \definecolor{LightSlateGray}{rgb}{0.466667,0.533333,0.600000}
  \definecolor{light slate grey}{rgb}{0.466667,0.533333,0.600000}
  \definecolor{LightSlateGrey}{rgb}{0.466667,0.533333,0.600000}
  \definecolor{gray}{rgb}{0.745098,0.745098,0.745098}
  \definecolor{grey}{rgb}{0.745098,0.745098,0.745098}
  \definecolor{light grey}{rgb}{0.827451,0.827451,0.827451}
  \definecolor{LightGrey}{rgb}{0.827451,0.827451,0.827451}
  \definecolor{light gray}{rgb}{0.827451,0.827451,0.827451}
  \definecolor{LightGray}{rgb}{0.827451,0.827451,0.827451}
  \definecolor{midnight blue}{rgb}{0.098039,0.098039,0.439216}
  \definecolor{MidnightBlue}{rgb}{0.098039,0.098039,0.439216}
  \definecolor{navy}{rgb}{0.000000,0.000000,0.501961}
  \definecolor{navy blue}{rgb}{0.000000,0.000000,0.501961}
  \definecolor{NavyBlue}{rgb}{0.000000,0.000000,0.501961}
  \definecolor{cornflower blue}{rgb}{0.392157,0.584314,0.929412}
  \definecolor{CornflowerBlue}{rgb}{0.392157,0.584314,0.929412}
  \definecolor{dark slate blue}{rgb}{0.282353,0.239216,0.545098}
  \definecolor{DarkSlateBlue}{rgb}{0.282353,0.239216,0.545098}
  \definecolor{slate blue}{rgb}{0.415686,0.352941,0.803922}
  \definecolor{SlateBlue}{rgb}{0.415686,0.352941,0.803922}
  \definecolor{medium slate blue}{rgb}{0.482353,0.407843,0.933333}
  \definecolor{MediumSlateBlue}{rgb}{0.482353,0.407843,0.933333}
  \definecolor{light slate blue}{rgb}{0.517647,0.439216,1.000000}
  \definecolor{LightSlateBlue}{rgb}{0.517647,0.439216,1.000000}
  \definecolor{medium blue}{rgb}{0.000000,0.000000,0.803922}
  \definecolor{MediumBlue}{rgb}{0.000000,0.000000,0.803922}
  \definecolor{royal blue}{rgb}{0.254902,0.411765,0.882353}
  \definecolor{RoyalBlue}{rgb}{0.254902,0.411765,0.882353}
  \definecolor{blue}{rgb}{0.000000,0.000000,1.000000}
  \definecolor{dodger blue}{rgb}{0.117647,0.564706,1.000000}
  \definecolor{DodgerBlue}{rgb}{0.117647,0.564706,1.000000}
  \definecolor{deep sky blue}{rgb}{0.000000,0.749020,1.000000}
  \definecolor{DeepSkyBlue}{rgb}{0.000000,0.749020,1.000000}
  \definecolor{sky blue}{rgb}{0.529412,0.807843,0.921569}
  \definecolor{SkyBlue}{rgb}{0.529412,0.807843,0.921569}
  \definecolor{light sky blue}{rgb}{0.529412,0.807843,0.980392}
  \definecolor{LightSkyBlue}{rgb}{0.529412,0.807843,0.980392}
  \definecolor{steel blue}{rgb}{0.274510,0.509804,0.705882}
  \definecolor{SteelBlue}{rgb}{0.274510,0.509804,0.705882}
  \definecolor{light steel blue}{rgb}{0.690196,0.768627,0.870588}
  \definecolor{LightSteelBlue}{rgb}{0.690196,0.768627,0.870588}
  \definecolor{light blue}{rgb}{0.678431,0.847059,0.901961}
  \definecolor{LightBlue}{rgb}{0.678431,0.847059,0.901961}
  \definecolor{powder blue}{rgb}{0.690196,0.878431,0.901961}
  \definecolor{PowderBlue}{rgb}{0.690196,0.878431,0.901961}
  \definecolor{pale turquoise}{rgb}{0.686275,0.933333,0.933333}
  \definecolor{PaleTurquoise}{rgb}{0.686275,0.933333,0.933333}
  \definecolor{dark turquoise}{rgb}{0.000000,0.807843,0.819608}
  \definecolor{DarkTurquoise}{rgb}{0.000000,0.807843,0.819608}
  \definecolor{medium turquoise}{rgb}{0.282353,0.819608,0.800000}
  \definecolor{MediumTurquoise}{rgb}{0.282353,0.819608,0.800000}
  \definecolor{turquoise}{rgb}{0.250980,0.878431,0.815686}
  \definecolor{cyan}{rgb}{0.000000,1.000000,1.000000}
  \definecolor{light cyan}{rgb}{0.878431,1.000000,1.000000}
  \definecolor{LightCyan}{rgb}{0.878431,1.000000,1.000000}
  \definecolor{cadet blue}{rgb}{0.372549,0.619608,0.627451}
  \definecolor{CadetBlue}{rgb}{0.372549,0.619608,0.627451}
  \definecolor{medium aquamarine}{rgb}{0.400000,0.803922,0.666667}
  \definecolor{MediumAquamarine}{rgb}{0.400000,0.803922,0.666667}
  \definecolor{aquamarine}{rgb}{0.498039,1.000000,0.831373}
  \definecolor{dark green}{rgb}{0.000000,0.392157,0.000000}
  \definecolor{DarkGreen}{rgb}{0.000000,0.392157,0.000000}
  \definecolor{dark olive green}{rgb}{0.333333,0.419608,0.184314}
  \definecolor{DarkOliveGreen}{rgb}{0.333333,0.419608,0.184314}
  \definecolor{dark sea green}{rgb}{0.560784,0.737255,0.560784}
  \definecolor{DarkSeaGreen}{rgb}{0.560784,0.737255,0.560784}
  \definecolor{sea green}{rgb}{0.180392,0.545098,0.341176}
  \definecolor{SeaGreen}{rgb}{0.180392,0.545098,0.341176}
  \definecolor{medium sea green}{rgb}{0.235294,0.701961,0.443137}
  \definecolor{MediumSeaGreen}{rgb}{0.235294,0.701961,0.443137}
  \definecolor{light sea green}{rgb}{0.125490,0.698039,0.666667}
  \definecolor{LightSeaGreen}{rgb}{0.125490,0.698039,0.666667}
  \definecolor{pale green}{rgb}{0.596078,0.984314,0.596078}
  \definecolor{PaleGreen}{rgb}{0.596078,0.984314,0.596078}
  \definecolor{spring green}{rgb}{0.000000,1.000000,0.498039}
  \definecolor{SpringGreen}{rgb}{0.000000,1.000000,0.498039}
  \definecolor{lawn green}{rgb}{0.486275,0.988235,0.000000}
  \definecolor{LawnGreen}{rgb}{0.486275,0.988235,0.000000}
  \definecolor{green}{rgb}{0.000000,1.000000,0.000000}
  \definecolor{chartreuse}{rgb}{0.498039,1.000000,0.000000}
  \definecolor{medium spring green}{rgb}{0.000000,0.980392,0.603922}
  \definecolor{MediumSpringGreen}{rgb}{0.000000,0.980392,0.603922}
  \definecolor{green yellow}{rgb}{0.678431,1.000000,0.184314}
  \definecolor{GreenYellow}{rgb}{0.678431,1.000000,0.184314}
  \definecolor{lime green}{rgb}{0.196078,0.803922,0.196078}
  \definecolor{LimeGreen}{rgb}{0.196078,0.803922,0.196078}
  \definecolor{yellow green}{rgb}{0.603922,0.803922,0.196078}
  \definecolor{YellowGreen}{rgb}{0.603922,0.803922,0.196078}
  \definecolor{forest green}{rgb}{0.133333,0.545098,0.133333}
  \definecolor{ForestGreen}{rgb}{0.133333,0.545098,0.133333}
  \definecolor{olive drab}{rgb}{0.419608,0.556863,0.137255}
  \definecolor{OliveDrab}{rgb}{0.419608,0.556863,0.137255}
  \definecolor{dark khaki}{rgb}{0.741176,0.717647,0.419608}
  \definecolor{DarkKhaki}{rgb}{0.741176,0.717647,0.419608}
  \definecolor{khaki}{rgb}{0.941176,0.901961,0.549020}
  \definecolor{pale goldenrod}{rgb}{0.933333,0.909804,0.666667}
  \definecolor{PaleGoldenrod}{rgb}{0.933333,0.909804,0.666667}
  \definecolor{light goldenrod yellow}{rgb}{0.980392,0.980392,0.823529}
  \definecolor{LightGoldenrodYellow}{rgb}{0.980392,0.980392,0.823529}
  \definecolor{light yellow}{rgb}{1.000000,1.000000,0.878431}
  \definecolor{LightYellow}{rgb}{1.000000,1.000000,0.878431}
  \definecolor{yellow}{rgb}{1.000000,1.000000,0.000000}
  \definecolor{gold}{rgb}{1.000000,0.843137,0.000000}
  \definecolor{light goldenrod}{rgb}{0.933333,0.866667,0.509804}
  \definecolor{LightGoldenrod}{rgb}{0.933333,0.866667,0.509804}
  \definecolor{goldenrod}{rgb}{0.854902,0.647059,0.125490}
  \definecolor{dark goldenrod}{rgb}{0.721569,0.525490,0.043137}
  \definecolor{DarkGoldenrod}{rgb}{0.721569,0.525490,0.043137}
  \definecolor{rosy brown}{rgb}{0.737255,0.560784,0.560784}
  \definecolor{RosyBrown}{rgb}{0.737255,0.560784,0.560784}
  \definecolor{indian red}{rgb}{0.803922,0.360784,0.360784}
  \definecolor{IndianRed}{rgb}{0.803922,0.360784,0.360784}
  \definecolor{saddle brown}{rgb}{0.545098,0.270588,0.074510}
  \definecolor{SaddleBrown}{rgb}{0.545098,0.270588,0.074510}
  \definecolor{sienna}{rgb}{0.627451,0.321569,0.176471}
  \definecolor{peru}{rgb}{0.803922,0.521569,0.247059}
  \definecolor{burlywood}{rgb}{0.870588,0.721569,0.529412}
  \definecolor{beige}{rgb}{0.960784,0.960784,0.862745}
  \definecolor{wheat}{rgb}{0.960784,0.870588,0.701961}
  \definecolor{sandy brown}{rgb}{0.956863,0.643137,0.376471}
  \definecolor{SandyBrown}{rgb}{0.956863,0.643137,0.376471}
  \definecolor{tan}{rgb}{0.823529,0.705882,0.549020}
  \definecolor{chocolate}{rgb}{0.823529,0.411765,0.117647}
  \definecolor{firebrick}{rgb}{0.698039,0.133333,0.133333}
  \definecolor{brown}{rgb}{0.647059,0.164706,0.164706}
  \definecolor{dark salmon}{rgb}{0.913725,0.588235,0.478431}
  \definecolor{DarkSalmon}{rgb}{0.913725,0.588235,0.478431}
  \definecolor{salmon}{rgb}{0.980392,0.501961,0.447059}
  \definecolor{light salmon}{rgb}{1.000000,0.627451,0.478431}
  \definecolor{LightSalmon}{rgb}{1.000000,0.627451,0.478431}
  \definecolor{orange}{rgb}{1.000000,0.647059,0.000000}
  \definecolor{dark orange}{rgb}{1.000000,0.549020,0.000000}
  \definecolor{DarkOrange}{rgb}{1.000000,0.549020,0.000000}
  \definecolor{coral}{rgb}{1.000000,0.498039,0.313726}
  \definecolor{light coral}{rgb}{0.941176,0.501961,0.501961}
  \definecolor{LightCoral}{rgb}{0.941176,0.501961,0.501961}
  \definecolor{tomato}{rgb}{1.000000,0.388235,0.278431}
  \definecolor{orange red}{rgb}{1.000000,0.270588,0.000000}
  \definecolor{OrangeRed}{rgb}{1.000000,0.270588,0.000000}
  \definecolor{red}{rgb}{1.000000,0.000000,0.000000}
  \definecolor{hot pink}{rgb}{1.000000,0.411765,0.705882}
  \definecolor{HotPink}{rgb}{1.000000,0.411765,0.705882}
  \definecolor{deep pink}{rgb}{1.000000,0.078431,0.576471}
  \definecolor{DeepPink}{rgb}{1.000000,0.078431,0.576471}
  \definecolor{pink}{rgb}{1.000000,0.752941,0.796078}
  \definecolor{light pink}{rgb}{1.000000,0.713726,0.756863}
  \definecolor{LightPink}{rgb}{1.000000,0.713726,0.756863}
  \definecolor{pale violet red}{rgb}{0.858824,0.439216,0.576471}
  \definecolor{PaleVioletRed}{rgb}{0.858824,0.439216,0.576471}
  \definecolor{maroon}{rgb}{0.690196,0.188235,0.376471}
  \definecolor{medium violet red}{rgb}{0.780392,0.082353,0.521569}
  \definecolor{MediumVioletRed}{rgb}{0.780392,0.082353,0.521569}
  \definecolor{violet red}{rgb}{0.815686,0.125490,0.564706}
  \definecolor{VioletRed}{rgb}{0.815686,0.125490,0.564706}
  \definecolor{magenta}{rgb}{1.000000,0.000000,1.000000}
  \definecolor{violet}{rgb}{0.933333,0.509804,0.933333}
  \definecolor{plum}{rgb}{0.866667,0.627451,0.866667}
  \definecolor{orchid}{rgb}{0.854902,0.439216,0.839216}
  \definecolor{medium orchid}{rgb}{0.729412,0.333333,0.827451}
  \definecolor{MediumOrchid}{rgb}{0.729412,0.333333,0.827451}
  \definecolor{dark orchid}{rgb}{0.600000,0.196078,0.800000}
  \definecolor{DarkOrchid}{rgb}{0.600000,0.196078,0.800000}
  \definecolor{dark violet}{rgb}{0.580392,0.000000,0.827451}
  \definecolor{DarkViolet}{rgb}{0.580392,0.000000,0.827451}
  \definecolor{blue violet}{rgb}{0.541176,0.168627,0.886275}
  \definecolor{BlueViolet}{rgb}{0.541176,0.168627,0.886275}
  \definecolor{purple}{rgb}{0.627451,0.125490,0.941176}
  \definecolor{medium purple}{rgb}{0.576471,0.439216,0.858824}
  \definecolor{MediumPurple}{rgb}{0.576471,0.439216,0.858824}
  \definecolor{thistle}{rgb}{0.847059,0.749020,0.847059}
  \definecolor{snow1}{rgb}{1.000000,0.980392,0.980392}
  \definecolor{snow2}{rgb}{0.933333,0.913725,0.913725}
  \definecolor{snow3}{rgb}{0.803922,0.788235,0.788235}
  \definecolor{snow4}{rgb}{0.545098,0.537255,0.537255}
  \definecolor{seashell1}{rgb}{1.000000,0.960784,0.933333}
  \definecolor{seashell2}{rgb}{0.933333,0.898039,0.870588}
  \definecolor{seashell3}{rgb}{0.803922,0.772549,0.749020}
  \definecolor{seashell4}{rgb}{0.545098,0.525490,0.509804}
  \definecolor{AntiqueWhite1}{rgb}{1.000000,0.937255,0.858824}
  \definecolor{AntiqueWhite2}{rgb}{0.933333,0.874510,0.800000}
  \definecolor{AntiqueWhite3}{rgb}{0.803922,0.752941,0.690196}
  \definecolor{AntiqueWhite4}{rgb}{0.545098,0.513726,0.470588}
  \definecolor{bisque1}{rgb}{1.000000,0.894118,0.768627}
  \definecolor{bisque2}{rgb}{0.933333,0.835294,0.717647}
  \definecolor{bisque3}{rgb}{0.803922,0.717647,0.619608}
  \definecolor{bisque4}{rgb}{0.545098,0.490196,0.419608}
  \definecolor{PeachPuff1}{rgb}{1.000000,0.854902,0.725490}
  \definecolor{PeachPuff2}{rgb}{0.933333,0.796078,0.678431}
  \definecolor{PeachPuff3}{rgb}{0.803922,0.686275,0.584314}
  \definecolor{PeachPuff4}{rgb}{0.545098,0.466667,0.396078}
  \definecolor{NavajoWhite1}{rgb}{1.000000,0.870588,0.678431}
  \definecolor{NavajoWhite2}{rgb}{0.933333,0.811765,0.631373}
  \definecolor{NavajoWhite3}{rgb}{0.803922,0.701961,0.545098}
  \definecolor{NavajoWhite4}{rgb}{0.545098,0.474510,0.368627}
  \definecolor{LemonChiffon1}{rgb}{1.000000,0.980392,0.803922}
  \definecolor{LemonChiffon2}{rgb}{0.933333,0.913725,0.749020}
  \definecolor{LemonChiffon3}{rgb}{0.803922,0.788235,0.647059}
  \definecolor{LemonChiffon4}{rgb}{0.545098,0.537255,0.439216}
  \definecolor{cornsilk1}{rgb}{1.000000,0.972549,0.862745}
  \definecolor{cornsilk2}{rgb}{0.933333,0.909804,0.803922}
  \definecolor{cornsilk3}{rgb}{0.803922,0.784314,0.694118}
  \definecolor{cornsilk4}{rgb}{0.545098,0.533333,0.470588}
  \definecolor{ivory1}{rgb}{1.000000,1.000000,0.941176}
  \definecolor{ivory2}{rgb}{0.933333,0.933333,0.878431}
  \definecolor{ivory3}{rgb}{0.803922,0.803922,0.756863}
  \definecolor{ivory4}{rgb}{0.545098,0.545098,0.513726}
  \definecolor{honeydew1}{rgb}{0.941176,1.000000,0.941176}
  \definecolor{honeydew2}{rgb}{0.878431,0.933333,0.878431}
  \definecolor{honeydew3}{rgb}{0.756863,0.803922,0.756863}
  \definecolor{honeydew4}{rgb}{0.513726,0.545098,0.513726}
  \definecolor{LavenderBlush1}{rgb}{1.000000,0.941176,0.960784}
  \definecolor{LavenderBlush2}{rgb}{0.933333,0.878431,0.898039}
  \definecolor{LavenderBlush3}{rgb}{0.803922,0.756863,0.772549}
  \definecolor{LavenderBlush4}{rgb}{0.545098,0.513726,0.525490}
  \definecolor{MistyRose1}{rgb}{1.000000,0.894118,0.882353}
  \definecolor{MistyRose2}{rgb}{0.933333,0.835294,0.823529}
  \definecolor{MistyRose3}{rgb}{0.803922,0.717647,0.709804}
  \definecolor{MistyRose4}{rgb}{0.545098,0.490196,0.482353}
  \definecolor{azure1}{rgb}{0.941176,1.000000,1.000000}
  \definecolor{azure2}{rgb}{0.878431,0.933333,0.933333}
  \definecolor{azure3}{rgb}{0.756863,0.803922,0.803922}
  \definecolor{azure4}{rgb}{0.513726,0.545098,0.545098}
  \definecolor{SlateBlue1}{rgb}{0.513726,0.435294,1.000000}
  \definecolor{SlateBlue2}{rgb}{0.478431,0.403922,0.933333}
  \definecolor{SlateBlue3}{rgb}{0.411765,0.349020,0.803922}
  \definecolor{SlateBlue4}{rgb}{0.278431,0.235294,0.545098}
  \definecolor{RoyalBlue1}{rgb}{0.282353,0.462745,1.000000}
  \definecolor{RoyalBlue2}{rgb}{0.262745,0.431373,0.933333}
  \definecolor{RoyalBlue3}{rgb}{0.227451,0.372549,0.803922}
  \definecolor{RoyalBlue4}{rgb}{0.152941,0.250980,0.545098}
  \definecolor{blue1}{rgb}{0.000000,0.000000,1.000000}
  \definecolor{blue2}{rgb}{0.000000,0.000000,0.933333}
  \definecolor{blue3}{rgb}{0.000000,0.000000,0.803922}
  \definecolor{blue4}{rgb}{0.000000,0.000000,0.545098}
  \definecolor{DodgerBlue1}{rgb}{0.117647,0.564706,1.000000}
  \definecolor{DodgerBlue2}{rgb}{0.109804,0.525490,0.933333}
  \definecolor{DodgerBlue3}{rgb}{0.094118,0.454902,0.803922}
  \definecolor{DodgerBlue4}{rgb}{0.062745,0.305882,0.545098}
  \definecolor{SteelBlue1}{rgb}{0.388235,0.721569,1.000000}
  \definecolor{SteelBlue2}{rgb}{0.360784,0.674510,0.933333}
  \definecolor{SteelBlue3}{rgb}{0.309804,0.580392,0.803922}
  \definecolor{SteelBlue4}{rgb}{0.211765,0.392157,0.545098}
  \definecolor{DeepSkyBlue1}{rgb}{0.000000,0.749020,1.000000}
  \definecolor{DeepSkyBlue2}{rgb}{0.000000,0.698039,0.933333}
  \definecolor{DeepSkyBlue3}{rgb}{0.000000,0.603922,0.803922}
  \definecolor{DeepSkyBlue4}{rgb}{0.000000,0.407843,0.545098}
  \definecolor{SkyBlue1}{rgb}{0.529412,0.807843,1.000000}
  \definecolor{SkyBlue2}{rgb}{0.494118,0.752941,0.933333}
  \definecolor{SkyBlue3}{rgb}{0.423529,0.650980,0.803922}
  \definecolor{SkyBlue4}{rgb}{0.290196,0.439216,0.545098}
  \definecolor{LightSkyBlue1}{rgb}{0.690196,0.886275,1.000000}
  \definecolor{LightSkyBlue2}{rgb}{0.643137,0.827451,0.933333}
  \definecolor{LightSkyBlue3}{rgb}{0.552941,0.713726,0.803922}
  \definecolor{LightSkyBlue4}{rgb}{0.376471,0.482353,0.545098}
  \definecolor{SlateGray1}{rgb}{0.776471,0.886275,1.000000}
  \definecolor{SlateGray2}{rgb}{0.725490,0.827451,0.933333}
  \definecolor{SlateGray3}{rgb}{0.623529,0.713726,0.803922}
  \definecolor{SlateGray4}{rgb}{0.423529,0.482353,0.545098}
  \definecolor{LightSteelBlue1}{rgb}{0.792157,0.882353,1.000000}
  \definecolor{LightSteelBlue2}{rgb}{0.737255,0.823529,0.933333}
  \definecolor{LightSteelBlue3}{rgb}{0.635294,0.709804,0.803922}
  \definecolor{LightSteelBlue4}{rgb}{0.431373,0.482353,0.545098}
  \definecolor{LightBlue1}{rgb}{0.749020,0.937255,1.000000}
  \definecolor{LightBlue2}{rgb}{0.698039,0.874510,0.933333}
  \definecolor{LightBlue3}{rgb}{0.603922,0.752941,0.803922}
  \definecolor{LightBlue4}{rgb}{0.407843,0.513726,0.545098}
  \definecolor{LightCyan1}{rgb}{0.878431,1.000000,1.000000}
  \definecolor{LightCyan2}{rgb}{0.819608,0.933333,0.933333}
  \definecolor{LightCyan3}{rgb}{0.705882,0.803922,0.803922}
  \definecolor{LightCyan4}{rgb}{0.478431,0.545098,0.545098}
  \definecolor{PaleTurquoise1}{rgb}{0.733333,1.000000,1.000000}
  \definecolor{PaleTurquoise2}{rgb}{0.682353,0.933333,0.933333}
  \definecolor{PaleTurquoise3}{rgb}{0.588235,0.803922,0.803922}
  \definecolor{PaleTurquoise4}{rgb}{0.400000,0.545098,0.545098}
  \definecolor{CadetBlue1}{rgb}{0.596078,0.960784,1.000000}
  \definecolor{CadetBlue2}{rgb}{0.556863,0.898039,0.933333}
  \definecolor{CadetBlue3}{rgb}{0.478431,0.772549,0.803922}
  \definecolor{CadetBlue4}{rgb}{0.325490,0.525490,0.545098}
  \definecolor{turquoise1}{rgb}{0.000000,0.960784,1.000000}
  \definecolor{turquoise2}{rgb}{0.000000,0.898039,0.933333}
  \definecolor{turquoise3}{rgb}{0.000000,0.772549,0.803922}
  \definecolor{turquoise4}{rgb}{0.000000,0.525490,0.545098}
  \definecolor{cyan1}{rgb}{0.000000,1.000000,1.000000}
  \definecolor{cyan2}{rgb}{0.000000,0.933333,0.933333}
  \definecolor{cyan3}{rgb}{0.000000,0.803922,0.803922}
  \definecolor{cyan4}{rgb}{0.000000,0.545098,0.545098}
  \definecolor{DarkSlateGray1}{rgb}{0.592157,1.000000,1.000000}
  \definecolor{DarkSlateGray2}{rgb}{0.552941,0.933333,0.933333}
  \definecolor{DarkSlateGray3}{rgb}{0.474510,0.803922,0.803922}
  \definecolor{DarkSlateGray4}{rgb}{0.321569,0.545098,0.545098}
  \definecolor{aquamarine1}{rgb}{0.498039,1.000000,0.831373}
  \definecolor{aquamarine2}{rgb}{0.462745,0.933333,0.776471}
  \definecolor{aquamarine3}{rgb}{0.400000,0.803922,0.666667}
  \definecolor{aquamarine4}{rgb}{0.270588,0.545098,0.454902}
  \definecolor{DarkSeaGreen1}{rgb}{0.756863,1.000000,0.756863}
  \definecolor{DarkSeaGreen2}{rgb}{0.705882,0.933333,0.705882}
  \definecolor{DarkSeaGreen3}{rgb}{0.607843,0.803922,0.607843}
  \definecolor{DarkSeaGreen4}{rgb}{0.411765,0.545098,0.411765}
  \definecolor{SeaGreen1}{rgb}{0.329412,1.000000,0.623529}
  \definecolor{SeaGreen2}{rgb}{0.305882,0.933333,0.580392}
  \definecolor{SeaGreen3}{rgb}{0.262745,0.803922,0.501961}
  \definecolor{SeaGreen4}{rgb}{0.180392,0.545098,0.341176}
  \definecolor{PaleGreen1}{rgb}{0.603922,1.000000,0.603922}
  \definecolor{PaleGreen2}{rgb}{0.564706,0.933333,0.564706}
  \definecolor{PaleGreen3}{rgb}{0.486275,0.803922,0.486275}
  \definecolor{PaleGreen4}{rgb}{0.329412,0.545098,0.329412}
  \definecolor{SpringGreen1}{rgb}{0.000000,1.000000,0.498039}
  \definecolor{SpringGreen2}{rgb}{0.000000,0.933333,0.462745}
  \definecolor{SpringGreen3}{rgb}{0.000000,0.803922,0.400000}
  \definecolor{SpringGreen4}{rgb}{0.000000,0.545098,0.270588}
  \definecolor{green1}{rgb}{0.000000,1.000000,0.000000}
  \definecolor{green2}{rgb}{0.000000,0.933333,0.000000}
  \definecolor{green3}{rgb}{0.000000,0.803922,0.000000}
  \definecolor{green4}{rgb}{0.000000,0.545098,0.000000}
  \definecolor{chartreuse1}{rgb}{0.498039,1.000000,0.000000}
  \definecolor{chartreuse2}{rgb}{0.462745,0.933333,0.000000}
  \definecolor{chartreuse3}{rgb}{0.400000,0.803922,0.000000}
  \definecolor{chartreuse4}{rgb}{0.270588,0.545098,0.000000}
  \definecolor{OliveDrab1}{rgb}{0.752941,1.000000,0.243137}
  \definecolor{OliveDrab2}{rgb}{0.701961,0.933333,0.227451}
  \definecolor{OliveDrab3}{rgb}{0.603922,0.803922,0.196078}
  \definecolor{OliveDrab4}{rgb}{0.411765,0.545098,0.133333}
  \definecolor{DarkOliveGreen1}{rgb}{0.792157,1.000000,0.439216}
  \definecolor{DarkOliveGreen2}{rgb}{0.737255,0.933333,0.407843}
  \definecolor{DarkOliveGreen3}{rgb}{0.635294,0.803922,0.352941}
  \definecolor{DarkOliveGreen4}{rgb}{0.431373,0.545098,0.239216}
  \definecolor{khaki1}{rgb}{1.000000,0.964706,0.560784}
  \definecolor{khaki2}{rgb}{0.933333,0.901961,0.521569}
  \definecolor{khaki3}{rgb}{0.803922,0.776471,0.450980}
  \definecolor{khaki4}{rgb}{0.545098,0.525490,0.305882}
  \definecolor{LightGoldenrod1}{rgb}{1.000000,0.925490,0.545098}
  \definecolor{LightGoldenrod2}{rgb}{0.933333,0.862745,0.509804}
  \definecolor{LightGoldenrod3}{rgb}{0.803922,0.745098,0.439216}
  \definecolor{LightGoldenrod4}{rgb}{0.545098,0.505882,0.298039}
  \definecolor{LightYellow1}{rgb}{1.000000,1.000000,0.878431}
  \definecolor{LightYellow2}{rgb}{0.933333,0.933333,0.819608}
  \definecolor{LightYellow3}{rgb}{0.803922,0.803922,0.705882}
  \definecolor{LightYellow4}{rgb}{0.545098,0.545098,0.478431}
  \definecolor{yellow1}{rgb}{1.000000,1.000000,0.000000}
  \definecolor{yellow2}{rgb}{0.933333,0.933333,0.000000}
  \definecolor{yellow3}{rgb}{0.803922,0.803922,0.000000}
  \definecolor{yellow4}{rgb}{0.545098,0.545098,0.000000}
  \definecolor{gold1}{rgb}{1.000000,0.843137,0.000000}
  \definecolor{gold2}{rgb}{0.933333,0.788235,0.000000}
  \definecolor{gold3}{rgb}{0.803922,0.678431,0.000000}
  \definecolor{gold4}{rgb}{0.545098,0.458824,0.000000}
  \definecolor{goldenrod1}{rgb}{1.000000,0.756863,0.145098}
  \definecolor{goldenrod2}{rgb}{0.933333,0.705882,0.133333}
  \definecolor{goldenrod3}{rgb}{0.803922,0.607843,0.113725}
  \definecolor{goldenrod4}{rgb}{0.545098,0.411765,0.078431}
  \definecolor{DarkGoldenrod1}{rgb}{1.000000,0.725490,0.058824}
  \definecolor{DarkGoldenrod2}{rgb}{0.933333,0.678431,0.054902}
  \definecolor{DarkGoldenrod3}{rgb}{0.803922,0.584314,0.047059}
  \definecolor{DarkGoldenrod4}{rgb}{0.545098,0.396078,0.031373}
  \definecolor{RosyBrown1}{rgb}{1.000000,0.756863,0.756863}
  \definecolor{RosyBrown2}{rgb}{0.933333,0.705882,0.705882}
  \definecolor{RosyBrown3}{rgb}{0.803922,0.607843,0.607843}
  \definecolor{RosyBrown4}{rgb}{0.545098,0.411765,0.411765}
  \definecolor{IndianRed1}{rgb}{1.000000,0.415686,0.415686}
  \definecolor{IndianRed2}{rgb}{0.933333,0.388235,0.388235}
  \definecolor{IndianRed3}{rgb}{0.803922,0.333333,0.333333}
  \definecolor{IndianRed4}{rgb}{0.545098,0.227451,0.227451}
  \definecolor{sienna1}{rgb}{1.000000,0.509804,0.278431}
  \definecolor{sienna2}{rgb}{0.933333,0.474510,0.258824}
  \definecolor{sienna3}{rgb}{0.803922,0.407843,0.223529}
  \definecolor{sienna4}{rgb}{0.545098,0.278431,0.149020}
  \definecolor{burlywood1}{rgb}{1.000000,0.827451,0.607843}
  \definecolor{burlywood2}{rgb}{0.933333,0.772549,0.568627}
  \definecolor{burlywood3}{rgb}{0.803922,0.666667,0.490196}
  \definecolor{burlywood4}{rgb}{0.545098,0.450980,0.333333}
  \definecolor{wheat1}{rgb}{1.000000,0.905882,0.729412}
  \definecolor{wheat2}{rgb}{0.933333,0.847059,0.682353}
  \definecolor{wheat3}{rgb}{0.803922,0.729412,0.588235}
  \definecolor{wheat4}{rgb}{0.545098,0.494118,0.400000}
  \definecolor{tan1}{rgb}{1.000000,0.647059,0.309804}
  \definecolor{tan2}{rgb}{0.933333,0.603922,0.286275}
  \definecolor{tan3}{rgb}{0.803922,0.521569,0.247059}
  \definecolor{tan4}{rgb}{0.545098,0.352941,0.168627}
  \definecolor{chocolate1}{rgb}{1.000000,0.498039,0.141176}
  \definecolor{chocolate2}{rgb}{0.933333,0.462745,0.129412}
  \definecolor{chocolate3}{rgb}{0.803922,0.400000,0.113725}
  \definecolor{chocolate4}{rgb}{0.545098,0.270588,0.074510}
  \definecolor{firebrick1}{rgb}{1.000000,0.188235,0.188235}
  \definecolor{firebrick2}{rgb}{0.933333,0.172549,0.172549}
  \definecolor{firebrick3}{rgb}{0.803922,0.149020,0.149020}
  \definecolor{firebrick4}{rgb}{0.545098,0.101961,0.101961}
  \definecolor{brown1}{rgb}{1.000000,0.250980,0.250980}
  \definecolor{brown2}{rgb}{0.933333,0.231373,0.231373}
  \definecolor{brown3}{rgb}{0.803922,0.200000,0.200000}
  \definecolor{brown4}{rgb}{0.545098,0.137255,0.137255}
  \definecolor{salmon1}{rgb}{1.000000,0.549020,0.411765}
  \definecolor{salmon2}{rgb}{0.933333,0.509804,0.384314}
  \definecolor{salmon3}{rgb}{0.803922,0.439216,0.329412}
  \definecolor{salmon4}{rgb}{0.545098,0.298039,0.223529}
  \definecolor{LightSalmon1}{rgb}{1.000000,0.627451,0.478431}
  \definecolor{LightSalmon2}{rgb}{0.933333,0.584314,0.447059}
  \definecolor{LightSalmon3}{rgb}{0.803922,0.505882,0.384314}
  \definecolor{LightSalmon4}{rgb}{0.545098,0.341176,0.258824}
  \definecolor{orange1}{rgb}{1.000000,0.647059,0.000000}
  \definecolor{orange2}{rgb}{0.933333,0.603922,0.000000}
  \definecolor{orange3}{rgb}{0.803922,0.521569,0.000000}
  \definecolor{orange4}{rgb}{0.545098,0.352941,0.000000}
  \definecolor{DarkOrange1}{rgb}{1.000000,0.498039,0.000000}
  \definecolor{DarkOrange2}{rgb}{0.933333,0.462745,0.000000}
  \definecolor{DarkOrange3}{rgb}{0.803922,0.400000,0.000000}
  \definecolor{DarkOrange4}{rgb}{0.545098,0.270588,0.000000}
  \definecolor{coral1}{rgb}{1.000000,0.447059,0.337255}
  \definecolor{coral2}{rgb}{0.933333,0.415686,0.313726}
  \definecolor{coral3}{rgb}{0.803922,0.356863,0.270588}
  \definecolor{coral4}{rgb}{0.545098,0.243137,0.184314}
  \definecolor{tomato1}{rgb}{1.000000,0.388235,0.278431}
  \definecolor{tomato2}{rgb}{0.933333,0.360784,0.258824}
  \definecolor{tomato3}{rgb}{0.803922,0.309804,0.223529}
  \definecolor{tomato4}{rgb}{0.545098,0.211765,0.149020}
  \definecolor{OrangeRed1}{rgb}{1.000000,0.270588,0.000000}
  \definecolor{OrangeRed2}{rgb}{0.933333,0.250980,0.000000}
  \definecolor{OrangeRed3}{rgb}{0.803922,0.215686,0.000000}
  \definecolor{OrangeRed4}{rgb}{0.545098,0.145098,0.000000}
  \definecolor{red1}{rgb}{1.000000,0.000000,0.000000}
  \definecolor{red2}{rgb}{0.933333,0.000000,0.000000}
  \definecolor{red3}{rgb}{0.803922,0.000000,0.000000}
  \definecolor{red4}{rgb}{0.545098,0.000000,0.000000}
  \definecolor{DeepPink1}{rgb}{1.000000,0.078431,0.576471}
  \definecolor{DeepPink2}{rgb}{0.933333,0.070588,0.537255}
  \definecolor{DeepPink3}{rgb}{0.803922,0.062745,0.462745}
  \definecolor{DeepPink4}{rgb}{0.545098,0.039216,0.313726}
  \definecolor{HotPink1}{rgb}{1.000000,0.431373,0.705882}
  \definecolor{HotPink2}{rgb}{0.933333,0.415686,0.654902}
  \definecolor{HotPink3}{rgb}{0.803922,0.376471,0.564706}
  \definecolor{HotPink4}{rgb}{0.545098,0.227451,0.384314}
  \definecolor{pink1}{rgb}{1.000000,0.709804,0.772549}
  \definecolor{pink2}{rgb}{0.933333,0.662745,0.721569}
  \definecolor{pink3}{rgb}{0.803922,0.568627,0.619608}
  \definecolor{pink4}{rgb}{0.545098,0.388235,0.423529}
  \definecolor{LightPink1}{rgb}{1.000000,0.682353,0.725490}
  \definecolor{LightPink2}{rgb}{0.933333,0.635294,0.678431}
  \definecolor{LightPink3}{rgb}{0.803922,0.549020,0.584314}
  \definecolor{LightPink4}{rgb}{0.545098,0.372549,0.396078}
  \definecolor{PaleVioletRed1}{rgb}{1.000000,0.509804,0.670588}
  \definecolor{PaleVioletRed2}{rgb}{0.933333,0.474510,0.623529}
  \definecolor{PaleVioletRed3}{rgb}{0.803922,0.407843,0.537255}
  \definecolor{PaleVioletRed4}{rgb}{0.545098,0.278431,0.364706}
  \definecolor{maroon1}{rgb}{1.000000,0.203922,0.701961}
  \definecolor{maroon2}{rgb}{0.933333,0.188235,0.654902}
  \definecolor{maroon3}{rgb}{0.803922,0.160784,0.564706}
  \definecolor{maroon4}{rgb}{0.545098,0.109804,0.384314}
  \definecolor{VioletRed1}{rgb}{1.000000,0.243137,0.588235}
  \definecolor{VioletRed2}{rgb}{0.933333,0.227451,0.549020}
  \definecolor{VioletRed3}{rgb}{0.803922,0.196078,0.470588}
  \definecolor{VioletRed4}{rgb}{0.545098,0.133333,0.321569}
  \definecolor{magenta1}{rgb}{1.000000,0.000000,1.000000}
  \definecolor{magenta2}{rgb}{0.933333,0.000000,0.933333}
  \definecolor{magenta3}{rgb}{0.803922,0.000000,0.803922}
  \definecolor{magenta4}{rgb}{0.545098,0.000000,0.545098}
  \definecolor{orchid1}{rgb}{1.000000,0.513726,0.980392}
  \definecolor{orchid2}{rgb}{0.933333,0.478431,0.913725}
  \definecolor{orchid3}{rgb}{0.803922,0.411765,0.788235}
  \definecolor{orchid4}{rgb}{0.545098,0.278431,0.537255}
  \definecolor{plum1}{rgb}{1.000000,0.733333,1.000000}
  \definecolor{plum2}{rgb}{0.933333,0.682353,0.933333}
  \definecolor{plum3}{rgb}{0.803922,0.588235,0.803922}
  \definecolor{plum4}{rgb}{0.545098,0.400000,0.545098}
  \definecolor{MediumOrchid1}{rgb}{0.878431,0.400000,1.000000}
  \definecolor{MediumOrchid2}{rgb}{0.819608,0.372549,0.933333}
  \definecolor{MediumOrchid3}{rgb}{0.705882,0.321569,0.803922}
  \definecolor{MediumOrchid4}{rgb}{0.478431,0.215686,0.545098}
  \definecolor{DarkOrchid1}{rgb}{0.749020,0.243137,1.000000}
  \definecolor{DarkOrchid2}{rgb}{0.698039,0.227451,0.933333}
  \definecolor{DarkOrchid3}{rgb}{0.603922,0.196078,0.803922}
  \definecolor{DarkOrchid4}{rgb}{0.407843,0.133333,0.545098}
  \definecolor{purple1}{rgb}{0.607843,0.188235,1.000000}
  \definecolor{purple2}{rgb}{0.568627,0.172549,0.933333}
  \definecolor{purple3}{rgb}{0.490196,0.149020,0.803922}
  \definecolor{purple4}{rgb}{0.333333,0.101961,0.545098}
  \definecolor{MediumPurple1}{rgb}{0.670588,0.509804,1.000000}
  \definecolor{MediumPurple2}{rgb}{0.623529,0.474510,0.933333}
  \definecolor{MediumPurple3}{rgb}{0.537255,0.407843,0.803922}
  \definecolor{MediumPurple4}{rgb}{0.364706,0.278431,0.545098}
  \definecolor{thistle1}{rgb}{1.000000,0.882353,1.000000}
  \definecolor{thistle2}{rgb}{0.933333,0.823529,0.933333}
  \definecolor{thistle3}{rgb}{0.803922,0.709804,0.803922}
  \definecolor{thistle4}{rgb}{0.545098,0.482353,0.545098}
  \definecolor{gray0}{rgb}{0.000000,0.000000,0.000000}
  \definecolor{grey0}{rgb}{0.000000,0.000000,0.000000}
  \definecolor{gray1}{rgb}{0.011765,0.011765,0.011765}
  \definecolor{grey1}{rgb}{0.011765,0.011765,0.011765}
  \definecolor{gray2}{rgb}{0.019608,0.019608,0.019608}
  \definecolor{grey2}{rgb}{0.019608,0.019608,0.019608}
  \definecolor{gray3}{rgb}{0.031373,0.031373,0.031373}
  \definecolor{grey3}{rgb}{0.031373,0.031373,0.031373}
  \definecolor{gray4}{rgb}{0.039216,0.039216,0.039216}
  \definecolor{grey4}{rgb}{0.039216,0.039216,0.039216}
  \definecolor{gray5}{rgb}{0.050980,0.050980,0.050980}
  \definecolor{grey5}{rgb}{0.050980,0.050980,0.050980}
  \definecolor{gray6}{rgb}{0.058824,0.058824,0.058824}
  \definecolor{grey6}{rgb}{0.058824,0.058824,0.058824}
  \definecolor{gray7}{rgb}{0.070588,0.070588,0.070588}
  \definecolor{grey7}{rgb}{0.070588,0.070588,0.070588}
  \definecolor{gray8}{rgb}{0.078431,0.078431,0.078431}
  \definecolor{grey8}{rgb}{0.078431,0.078431,0.078431}
  \definecolor{gray9}{rgb}{0.090196,0.090196,0.090196}
  \definecolor{grey9}{rgb}{0.090196,0.090196,0.090196}
  \definecolor{gray10}{rgb}{0.101961,0.101961,0.101961}
  \definecolor{grey10}{rgb}{0.101961,0.101961,0.101961}
  \definecolor{gray11}{rgb}{0.109804,0.109804,0.109804}
  \definecolor{grey11}{rgb}{0.109804,0.109804,0.109804}
  \definecolor{gray12}{rgb}{0.121569,0.121569,0.121569}
  \definecolor{grey12}{rgb}{0.121569,0.121569,0.121569}
  \definecolor{gray13}{rgb}{0.129412,0.129412,0.129412}
  \definecolor{grey13}{rgb}{0.129412,0.129412,0.129412}
  \definecolor{gray14}{rgb}{0.141176,0.141176,0.141176}
  \definecolor{grey14}{rgb}{0.141176,0.141176,0.141176}
  \definecolor{gray15}{rgb}{0.149020,0.149020,0.149020}
  \definecolor{grey15}{rgb}{0.149020,0.149020,0.149020}
  \definecolor{gray16}{rgb}{0.160784,0.160784,0.160784}
  \definecolor{grey16}{rgb}{0.160784,0.160784,0.160784}
  \definecolor{gray17}{rgb}{0.168627,0.168627,0.168627}
  \definecolor{grey17}{rgb}{0.168627,0.168627,0.168627}
  \definecolor{gray18}{rgb}{0.180392,0.180392,0.180392}
  \definecolor{grey18}{rgb}{0.180392,0.180392,0.180392}
  \definecolor{gray19}{rgb}{0.188235,0.188235,0.188235}
  \definecolor{grey19}{rgb}{0.188235,0.188235,0.188235}
  \definecolor{gray20}{rgb}{0.200000,0.200000,0.200000}
  \definecolor{grey20}{rgb}{0.200000,0.200000,0.200000}
  \definecolor{gray21}{rgb}{0.211765,0.211765,0.211765}
  \definecolor{grey21}{rgb}{0.211765,0.211765,0.211765}
  \definecolor{gray22}{rgb}{0.219608,0.219608,0.219608}
  \definecolor{grey22}{rgb}{0.219608,0.219608,0.219608}
  \definecolor{gray23}{rgb}{0.231373,0.231373,0.231373}
  \definecolor{grey23}{rgb}{0.231373,0.231373,0.231373}
  \definecolor{gray24}{rgb}{0.239216,0.239216,0.239216}
  \definecolor{grey24}{rgb}{0.239216,0.239216,0.239216}
  \definecolor{gray25}{rgb}{0.250980,0.250980,0.250980}
  \definecolor{grey25}{rgb}{0.250980,0.250980,0.250980}
  \definecolor{gray26}{rgb}{0.258824,0.258824,0.258824}
  \definecolor{grey26}{rgb}{0.258824,0.258824,0.258824}
  \definecolor{gray27}{rgb}{0.270588,0.270588,0.270588}
  \definecolor{grey27}{rgb}{0.270588,0.270588,0.270588}
  \definecolor{gray28}{rgb}{0.278431,0.278431,0.278431}
  \definecolor{grey28}{rgb}{0.278431,0.278431,0.278431}
  \definecolor{gray29}{rgb}{0.290196,0.290196,0.290196}
  \definecolor{grey29}{rgb}{0.290196,0.290196,0.290196}
  \definecolor{gray30}{rgb}{0.301961,0.301961,0.301961}
  \definecolor{grey30}{rgb}{0.301961,0.301961,0.301961}
  \definecolor{gray31}{rgb}{0.309804,0.309804,0.309804}
  \definecolor{grey31}{rgb}{0.309804,0.309804,0.309804}
  \definecolor{gray32}{rgb}{0.321569,0.321569,0.321569}
  \definecolor{grey32}{rgb}{0.321569,0.321569,0.321569}
  \definecolor{gray33}{rgb}{0.329412,0.329412,0.329412}
  \definecolor{grey33}{rgb}{0.329412,0.329412,0.329412}
  \definecolor{gray34}{rgb}{0.341176,0.341176,0.341176}
  \definecolor{grey34}{rgb}{0.341176,0.341176,0.341176}
  \definecolor{gray35}{rgb}{0.349020,0.349020,0.349020}
  \definecolor{grey35}{rgb}{0.349020,0.349020,0.349020}
  \definecolor{gray36}{rgb}{0.360784,0.360784,0.360784}
  \definecolor{grey36}{rgb}{0.360784,0.360784,0.360784}
  \definecolor{gray37}{rgb}{0.368627,0.368627,0.368627}
  \definecolor{grey37}{rgb}{0.368627,0.368627,0.368627}
  \definecolor{gray38}{rgb}{0.380392,0.380392,0.380392}
  \definecolor{grey38}{rgb}{0.380392,0.380392,0.380392}
  \definecolor{gray39}{rgb}{0.388235,0.388235,0.388235}
  \definecolor{grey39}{rgb}{0.388235,0.388235,0.388235}
  \definecolor{gray40}{rgb}{0.400000,0.400000,0.400000}
  \definecolor{grey40}{rgb}{0.400000,0.400000,0.400000}
  \definecolor{gray41}{rgb}{0.411765,0.411765,0.411765}
  \definecolor{grey41}{rgb}{0.411765,0.411765,0.411765}
  \definecolor{gray42}{rgb}{0.419608,0.419608,0.419608}
  \definecolor{grey42}{rgb}{0.419608,0.419608,0.419608}
  \definecolor{gray43}{rgb}{0.431373,0.431373,0.431373}
  \definecolor{grey43}{rgb}{0.431373,0.431373,0.431373}
  \definecolor{gray44}{rgb}{0.439216,0.439216,0.439216}
  \definecolor{grey44}{rgb}{0.439216,0.439216,0.439216}
  \definecolor{gray45}{rgb}{0.450980,0.450980,0.450980}
  \definecolor{grey45}{rgb}{0.450980,0.450980,0.450980}
  \definecolor{gray46}{rgb}{0.458824,0.458824,0.458824}
  \definecolor{grey46}{rgb}{0.458824,0.458824,0.458824}
  \definecolor{gray47}{rgb}{0.470588,0.470588,0.470588}
  \definecolor{grey47}{rgb}{0.470588,0.470588,0.470588}
  \definecolor{gray48}{rgb}{0.478431,0.478431,0.478431}
  \definecolor{grey48}{rgb}{0.478431,0.478431,0.478431}
  \definecolor{gray49}{rgb}{0.490196,0.490196,0.490196}
  \definecolor{grey49}{rgb}{0.490196,0.490196,0.490196}
  \definecolor{gray50}{rgb}{0.498039,0.498039,0.498039}
  \definecolor{grey50}{rgb}{0.498039,0.498039,0.498039}
  \definecolor{gray51}{rgb}{0.509804,0.509804,0.509804}
  \definecolor{grey51}{rgb}{0.509804,0.509804,0.509804}
  \definecolor{gray52}{rgb}{0.521569,0.521569,0.521569}
  \definecolor{grey52}{rgb}{0.521569,0.521569,0.521569}
  \definecolor{gray53}{rgb}{0.529412,0.529412,0.529412}
  \definecolor{grey53}{rgb}{0.529412,0.529412,0.529412}
  \definecolor{gray54}{rgb}{0.541176,0.541176,0.541176}
  \definecolor{grey54}{rgb}{0.541176,0.541176,0.541176}
  \definecolor{gray55}{rgb}{0.549020,0.549020,0.549020}
  \definecolor{grey55}{rgb}{0.549020,0.549020,0.549020}
  \definecolor{gray56}{rgb}{0.560784,0.560784,0.560784}
  \definecolor{grey56}{rgb}{0.560784,0.560784,0.560784}
  \definecolor{gray57}{rgb}{0.568627,0.568627,0.568627}
  \definecolor{grey57}{rgb}{0.568627,0.568627,0.568627}
  \definecolor{gray58}{rgb}{0.580392,0.580392,0.580392}
  \definecolor{grey58}{rgb}{0.580392,0.580392,0.580392}
  \definecolor{gray59}{rgb}{0.588235,0.588235,0.588235}
  \definecolor{grey59}{rgb}{0.588235,0.588235,0.588235}
  \definecolor{gray60}{rgb}{0.600000,0.600000,0.600000}
  \definecolor{grey60}{rgb}{0.600000,0.600000,0.600000}
  \definecolor{gray61}{rgb}{0.611765,0.611765,0.611765}
  \definecolor{grey61}{rgb}{0.611765,0.611765,0.611765}
  \definecolor{gray62}{rgb}{0.619608,0.619608,0.619608}
  \definecolor{grey62}{rgb}{0.619608,0.619608,0.619608}
  \definecolor{gray63}{rgb}{0.631373,0.631373,0.631373}
  \definecolor{grey63}{rgb}{0.631373,0.631373,0.631373}
  \definecolor{gray64}{rgb}{0.639216,0.639216,0.639216}
  \definecolor{grey64}{rgb}{0.639216,0.639216,0.639216}
  \definecolor{gray65}{rgb}{0.650980,0.650980,0.650980}
  \definecolor{grey65}{rgb}{0.650980,0.650980,0.650980}
  \definecolor{gray66}{rgb}{0.658824,0.658824,0.658824}
  \definecolor{grey66}{rgb}{0.658824,0.658824,0.658824}
  \definecolor{gray67}{rgb}{0.670588,0.670588,0.670588}
  \definecolor{grey67}{rgb}{0.670588,0.670588,0.670588}
  \definecolor{gray68}{rgb}{0.678431,0.678431,0.678431}
  \definecolor{grey68}{rgb}{0.678431,0.678431,0.678431}
  \definecolor{gray69}{rgb}{0.690196,0.690196,0.690196}
  \definecolor{grey69}{rgb}{0.690196,0.690196,0.690196}
  \definecolor{gray70}{rgb}{0.701961,0.701961,0.701961}
  \definecolor{grey70}{rgb}{0.701961,0.701961,0.701961}
  \definecolor{gray71}{rgb}{0.709804,0.709804,0.709804}
  \definecolor{grey71}{rgb}{0.709804,0.709804,0.709804}
  \definecolor{gray72}{rgb}{0.721569,0.721569,0.721569}
  \definecolor{grey72}{rgb}{0.721569,0.721569,0.721569}
  \definecolor{gray73}{rgb}{0.729412,0.729412,0.729412}
  \definecolor{grey73}{rgb}{0.729412,0.729412,0.729412}
  \definecolor{gray74}{rgb}{0.741176,0.741176,0.741176}
  \definecolor{grey74}{rgb}{0.741176,0.741176,0.741176}
  \definecolor{gray75}{rgb}{0.749020,0.749020,0.749020}
  \definecolor{grey75}{rgb}{0.749020,0.749020,0.749020}
  \definecolor{gray76}{rgb}{0.760784,0.760784,0.760784}
  \definecolor{grey76}{rgb}{0.760784,0.760784,0.760784}
  \definecolor{gray77}{rgb}{0.768627,0.768627,0.768627}
  \definecolor{grey77}{rgb}{0.768627,0.768627,0.768627}
  \definecolor{gray78}{rgb}{0.780392,0.780392,0.780392}
  \definecolor{grey78}{rgb}{0.780392,0.780392,0.780392}
  \definecolor{gray79}{rgb}{0.788235,0.788235,0.788235}
  \definecolor{grey79}{rgb}{0.788235,0.788235,0.788235}
  \definecolor{gray80}{rgb}{0.800000,0.800000,0.800000}
  \definecolor{grey80}{rgb}{0.800000,0.800000,0.800000}
  \definecolor{gray81}{rgb}{0.811765,0.811765,0.811765}
  \definecolor{grey81}{rgb}{0.811765,0.811765,0.811765}
  \definecolor{gray82}{rgb}{0.819608,0.819608,0.819608}
  \definecolor{grey82}{rgb}{0.819608,0.819608,0.819608}
  \definecolor{gray83}{rgb}{0.831373,0.831373,0.831373}
  \definecolor{grey83}{rgb}{0.831373,0.831373,0.831373}
  \definecolor{gray84}{rgb}{0.839216,0.839216,0.839216}
  \definecolor{grey84}{rgb}{0.839216,0.839216,0.839216}
  \definecolor{gray85}{rgb}{0.850980,0.850980,0.850980}
  \definecolor{grey85}{rgb}{0.850980,0.850980,0.850980}
  \definecolor{gray86}{rgb}{0.858824,0.858824,0.858824}
  \definecolor{grey86}{rgb}{0.858824,0.858824,0.858824}
  \definecolor{gray87}{rgb}{0.870588,0.870588,0.870588}
  \definecolor{grey87}{rgb}{0.870588,0.870588,0.870588}
  \definecolor{gray88}{rgb}{0.878431,0.878431,0.878431}
  \definecolor{grey88}{rgb}{0.878431,0.878431,0.878431}
  \definecolor{gray89}{rgb}{0.890196,0.890196,0.890196}
  \definecolor{grey89}{rgb}{0.890196,0.890196,0.890196}
  \definecolor{gray90}{rgb}{0.898039,0.898039,0.898039}
  \definecolor{grey90}{rgb}{0.898039,0.898039,0.898039}
  \definecolor{gray91}{rgb}{0.909804,0.909804,0.909804}
  \definecolor{grey91}{rgb}{0.909804,0.909804,0.909804}
  \definecolor{gray92}{rgb}{0.921569,0.921569,0.921569}
  \definecolor{grey92}{rgb}{0.921569,0.921569,0.921569}
  \definecolor{gray93}{rgb}{0.929412,0.929412,0.929412}
  \definecolor{grey93}{rgb}{0.929412,0.929412,0.929412}
  \definecolor{gray94}{rgb}{0.941176,0.941176,0.941176}
  \definecolor{grey94}{rgb}{0.941176,0.941176,0.941176}
  \definecolor{gray95}{rgb}{0.949020,0.949020,0.949020}
  \definecolor{grey95}{rgb}{0.949020,0.949020,0.949020}
  \definecolor{gray96}{rgb}{0.960784,0.960784,0.960784}
  \definecolor{grey96}{rgb}{0.960784,0.960784,0.960784}
  \definecolor{gray97}{rgb}{0.968627,0.968627,0.968627}
  \definecolor{grey97}{rgb}{0.968627,0.968627,0.968627}
  \definecolor{gray98}{rgb}{0.980392,0.980392,0.980392}
  \definecolor{grey98}{rgb}{0.980392,0.980392,0.980392}
  \definecolor{gray99}{rgb}{0.988235,0.988235,0.988235}
  \definecolor{grey99}{rgb}{0.988235,0.988235,0.988235}
  \definecolor{gray100}{rgb}{1.000000,1.000000,1.000000}
  \definecolor{grey100}{rgb}{1.000000,1.000000,1.000000}
  \definecolor{dark grey}{rgb}{0.662745,0.662745,0.662745}
  \definecolor{DarkGrey}{rgb}{0.662745,0.662745,0.662745}
  \definecolor{dark gray}{rgb}{0.662745,0.662745,0.662745}
  \definecolor{DarkGray}{rgb}{0.662745,0.662745,0.662745}
  \definecolor{dark blue}{rgb}{0.000000,0.000000,0.545098}
  \definecolor{DarkBlue}{rgb}{0.000000,0.000000,0.545098}
  \definecolor{dark cyan}{rgb}{0.000000,0.545098,0.545098}
  \definecolor{DarkCyan}{rgb}{0.000000,0.545098,0.545098}
  \definecolor{dark magenta}{rgb}{0.545098,0.000000,0.545098}
  \definecolor{DarkMagenta}{rgb}{0.545098,0.000000,0.545098}
  \definecolor{dark red}{rgb}{0.545098,0.000000,0.000000}
  \definecolor{DarkRed}{rgb}{0.545098,0.000000,0.000000}
  \definecolor{light green}{rgb}{0.564706,0.933333,0.564706}
  \definecolor{LightGreen}{rgb}{0.564706,0.933333,0.564706}

\def\ROSSO#1{{\color{red}#1}}
\def\BLU#1{{\color{blue}#1}}

\definecolor{ian}{RGB}{153,102,255}

\begin{document}
\centerline{
\BLU{\Large\it Non-equilibrium statistical mechanics of turbulence}}

\*
\centerline{\ROSSO{\bf Comments on Ruelle's intermittency theory}}
\centerline{\BLU{\rm Giovanni Gallavotti and Pedro Garrido}}
\*

\def\SEC{A hierarchical turbulence model}
\section{\SEC}
\iniz

 The proposal \cite{Ru012,Ru014} for a theory of the corrections to the OK
 theory (``intermittency corrections'') is to take into account that the
 Kolmogorov scale itsef should be regarded as a fluctuating
   variable.

The OK theory is implied by the assumption, for $n$ large, of zero average
work due to interactions between wave components with wave length
$<\k^{-n}\ell_0\equiv\ell_n$ and components with wave length
$>\k\,\k^{-n}\ell_0$ ($\ell_0$ being the length scale where the
energy is input in the fluid and $\k$ a scale factor to be determined)
together with the assumption of independence of the distribution of the
components with inverse wave length (``momentum'') in the shell
$[\k^n,\k\k^n]\ell_0^{-1}$, \cite[p.420]{Ga002}.

It is represented by the equalities
\be\frac{\V v_{ni}^3}{\ell_n}=\frac{\V v_{(n+1)i'}^3}{\ell_{n+1}},\qquad\V
v=|v|, \ v\in R^3\Eq{e1.1}\ee
interpreted as stating an {\it equality up to fluctuations} of the velocity
components of scale $\k^{-n}\ell_0$, \ie of the part of the velocity field
which can be represented by the Fourier components in a basis of plane
waves localized in boxes, labeled by $i=1,\ldots \k^{3n}$, of size
$\k^{-n}\ell_0$ into which the fluid (moving in a container of linear size
$\ell_0$) is imagined decomposed (a wavelet representation) so that
$(n+1,i')$ labels a box contained in the box $(n,i)$.

The length scales are supposed to be separated by a suitably large scale
factor $\k$ (\ie $\ell_n=\k^{-n}\ell_0=\k^{-1}\ell_{n-1}$) so that the
fluctuations can be considered independent, however not so large that more
than one scalar quantity (namely $\V v_{n,i}^3$) suffices to describe the
independent components of the velocity (small enough to avoid that
``several different temperatures will be present among the systems $(n + 1,
j')$'' inside the containing box labeled $(n, j)$, and the $\V v_{n+1,j}$
distribution ``will not be Boltzmannian for a constant
temperature inside'',\cite[p.2]{Ru012}).

The distribution of $\V v_{n+1,j}^3$ is then simply chosen so that the
average of the $\V v_{n+1,j}^3$ is the value $\V v_{n,i}^3\k$ if the $\V
v_{n+1,j}^3$ on scale $n+1$ gives a finer description of the field in a box
named $j$ contained in the box named $i$ of scale larger by one unit.

Among the distributions with this property is selected the one which
maximizes entropy\footnote{\small If the box $\D=(n,j)\subset \D'=(n-1,j')$
  then the distribution $\P(W|W_{\D'})$ of $W_\D\equiv \V v^3_\D$ is
  conditioned to be such that $\media{W}=\k^{-1} W_{\D'}$; therefore the
  maximum entropy condition is that $-\int \P(W|W')\log \P(W|W') dW-\l_\D
\int W \P(W|W')dW$, where $\l_\D$ is a Lagrange multiplier, is maximal under
the constraint that $\media{W}=W'\k^{-1}$: this gives the
expression, called {\it Boltzmannian} in \cite{Ru012}, for $\P(W|W')$.}
and is:
\be\eqalign{& W_{ni}\defi |v_{ni}|^3,\qquad
\prod_{m=0}^n \prod_{i=1}^{\k^m}\frac{ d W_{i,m+1} }{W_{i'm}}\k\,
e^{-\k\frac{W_{i,m+1}}{W_{i',m}}}\cr}\Eq{e1.2}\ee
with $W_0$ a constant that parameterizes the fixed energy input at large
scale: the motion will be supposed to have a $0$ average total velocity at
each point; hence $W_0^{\frac13}$ can be viewed as an imposed average
velocity gradient at the largest scale $\ell_0$.

The $\V v_{in}=W^{\frac13}_{in}$ is then interpreted as a velocity
variation on a box of scale $\ell_0\k^{-n}$ or $\k^{-n}$ as $\ell_0$ will
be taken $1$. The index $i$ will be often omitted as we shall mostly be
concerned about a chain of boxes, one per each scale
$\k^{-n},n=0,1,\ldots$, totally ordered by inclusion (\ie the box labeled
$(i,n)$ contains the box labeled $(i',n+1)$).

The distribution of the energy dissipation $W_{n,i}\defi \V v_{n,i}^3$ in
the hierarchically arranged sequence of cells is therefore close in spirit
to the hierarchical models that have been source of ideas and so much
impact, at the birth of the {\it renormalization group} approach to
multiscale phenomena, in quantum field theory, critical point statistical
mechanics, low temperature physics, Fourier series convergence to name a
few, and to their nonperturbative analysis, either phenomenological or
mathematically rigorous, \cite{Wi965,Dy969,Wi970,Wi975,BS973,Ga979,BG995}.

The present turbulent fluctuations model can
therefore be called {\it hierarchical model for turbulence}
in the inertial scales.
It will be supposed to describe the velocity fluctuations
at scales $n$ at which the Reynolds number is larger than $1$,
\ie as long as $\frac{\V v_n\k^{-n}\ell_0}\n>1$.

The description will of course be approximate, \cite[Sec.3]{Ru014}: for
instance the correlations of the velocity gradient components are not
considered (and skewness will still rely on the classic OK theory,
\cite[Sec.34]{LL987}).

Given the distribution (and the initial parameter $W_0$) it ``only''
remains to study its properties assuming the distribution valid for
velocity profiles such that $\V v_n\k^{-n}\ell_0>\n$ after
fixing the value of  $\k$ in order to
match data in the literature (as explained in \cite[Eq.(12)]{Ru014}). As a
first remark the scaling corrections proposed in [12] can be rederived.

The average energy dissipation in a box of scale $n$ can be defined as the
average of $\e_n\defi W_n\ell^{-n}, \, \ell_n=\ell_0\k^{-n}$: the latter
average and its $p$''th order moments
can be readily computed to be, for $p>0$:
\be\eqalign{
&\frac{\log\media{\e_n^p}}{-\log \ell_n}\tende{n\to\infty}
\t_p=-\frac{\log \G(1+p)}{\log\k},\qquad
\media{\e_n^p}\sim \k^{n\t_p},\cr
&\media{(\frac{W_n}{\ell_n})^{\frac{p}3}} \sim
\k^{n\t_{\frac{p}3}},\qquad
\media{\V v_n^p}\sim \ell_0^{\frac{p}3}\k^{-n\z_p},
\qquad\z_p=\frac{p}3+\t_p\cr}\Eq{e1.3}\ee
The $W^{\frac13}_n$ being interpreted as a velocity
variation on a box of
scale $\ell_0\k^{-n}$, the last formula can also be
read as exressing the
$\media{(\frac{|\D_r v|}r)^p}\sim r^{\z_p}$ with
$\z_p=\frac13-\frac{\log\G(\frac{p}3+1)}{\log\k}$.

The $\t_p$ is the intermittency correction to the value $\frac13$: the
latter is the standard value of the OK theory in which there is no
fluctuation of the dissipation per unit time and volume
$\frac{W_n}{\ell_n}$; this gives us one free parameter, namely $\k$, to fit
experimental data: its value, universal within Ruelle's theory, turns out
to be quite large, $\k\sim 22.75$, \cite{Ru012}, fitting quite well all
experimental $p$-values ($p<18$).

Other universal predictions are possible. In \cite{Ru014} a quantity has
been studied for which accurate simulations are available.

If $\V W$ is a sample $(W_0,W_1,\ldots)$ of
the dissipations at scales $0,1,\ldots$ for the distribution in the
hierarchical turbulence model, the smallest scale $n(\V W)$ at which
$W_n^{\frac13}\ell_0 \k^{-n}\simeq\n$ occurs is the scale at which the
Kolmogorv scale is attained (\ie the Reynolds number
$\frac{W_n^{\frac13}\ell_n}\n$ becomes $<1$).

Taking $\ell_0=1,\n=1$, at such (random) Kolmogorov scale the actual
dissipation is $\x=W_{n(\V W)}\k^{n(\V W)}$ with a probability distribution
with density $P^*(\x)$. If $w_k=\frac{W_k}{W_{k-1}}$ then $W_n=W_0w_1\cdots
  w_n$ and the computation of $P^*(\x)$ can be seen as a problem on extreme
  events about the value of a product of random variables. Hence is is
  natural that the analysis of $P^*$ involves the Gumbel distribution
  $\phi(t)$ (which appears  with parameter $3$), \cite{Ru014}.

The $P^*$ is a distribution (universal once the value of $\k$ has been
fixed to fit the mentioned intermittency data) which is interesting because
it can be related to a quantity studied in simulations.

It has been remarked, \cite{Ru014}, that, assuming a symmetric distribution
of the velocity increments on scale $\k^{-n}$ whose modulus is
$W_{n,i}^{\frac13}$, the hierarchical turbulence model can be applied to
study the distribution of the velocity increments: for small velocity
increments the calculation can be performed very explictly and
quantitatively precise results are derived, that can be conceivably checked
at least in simulations. The data analysis and the (straightfoward)
numerical evaluation of the distribution $P^*$ is described below,
following \cite{Ru014}.

\def\SEC{Data settings}
\section{\SEC}
\iniz

\0Let $\ell_0,\n=1$ and let $\V W=(W_0,W_1,\ldots)$ be a sample chosen with
the distribution

\be p(d \V W)=\prod_{i=1}^\infty
\frac{\k \,dW_i}{W_{i-1}}\, e^{-\k\frac{W_i}{W_{i-1}}}
\Eq{e2.1}\ee
with $W_0,\k$ given parameters; and let $\V
v=(v_0,v_1,\ldots)=(W_0^{\frac13},W_1^{\frac13},\ldots)$.

Define $n(\V W)=n$ as the smallest value of $i$ such that
$W_i^{\frac13}\k^{-i}\equiv v_i\k^{-i}<1$: $n(\V W)$ will be called the
``dissipation scale'' of $\V W$.  \*

Imagine to have a large number $\NN$ of $p$-distributed samples of $\V W$'s.
Given $h>0$ let
\be P_n^*(\x)\defi \frac1h \frac1\NN\Big( (\#\, \V W\, {\rm with}\, n(\V
W)=n)\cap
(\x<(W_n /W_0)^{\frac13}\k^n< \x+h)\Big)\Eq{e2.2}\ee
hence $h P^*_n(\x)$ is the probability that the dissipation scale $n$
is reached  with $\x$ in $[\x,\x+h]$. Then
$P^*(\x)\defi\sum_{n=0}^\infty P^*_n(\x)$
is the probability density that, at the dissipation scale, the velocity
gradient $\frac{v_n}{v_0}\k^n$ is between $\x$ and $\x+h$.

The velocity component in a direction is $v_n\cos\th$: so that the
probability that it is in $d\x$ with gradient $\frac{v_n}{v_0}\k^n$ and
that this happens at dissipation scale
$=n$ is $d\x$ times
\be \int P^*_n(\frac{v_n}{v_0}
\k^n=\x_0)d\x_0\d(\x_0\vert\cos\th\vert-\x)\frac{\sin\th d\th d\f}{4\p}= 
\int_{\x}^\infty \frac{P_n^*(\x_0)}{\x_0}d\x_0\Eq{e2.3}\ee
Let
\be
P(\x)\defi\int_{\x_0>\x}
\frac{d\x_0}{\x_0}\sum_{n=1}^\infty P^*_n(\x_0)
\Eq{e2.4}\ee
that is the probability distribution of the (normalized radial velocity
gradient) and
\be\s_m=\int_0^\infty d\x P(\x)\x^m
\Eq{e2.5}\ee
its momenta. To compare this distribution to experimental data
\cite{SSKDYS014} it is convenient to define
\be
p(z)=\frac{1}{2}\s_2^{1/2} P(\s_2^{1/2}\vert z\vert)\Eq{e2.6}
\ee

We have used the following computational algorithnm to $P(\xi)$:

\begin{itemize}
\item (1) Build a sample $(i)$ $\V W^{(i)}=(W_0,W_1,\ldots,W_n,\ldots)$
\item (2) Stop when $n=\bar n_i$ such that $W_{n-1}^{1/3}\kappa^{-(n-1)}>1>W_{n}^{1/3}\kappa^{-n}$
\item (3) Evaluate $\bar m_i= int(\xi_i/h)+1$ where  $\xi_i=\kappa^{\bar n_i}(W_{\bar n_i}/W_0)^{1/3}$ 
\item (4) goto to (1) during $N$ times

\end{itemize} 

Then, the distribution $P(\xi)$ is given by
\be
P(mh-h/2)=h^{-1}\bar P(m)\quad,\quad \bar P(m)=\frac{1}{N}\sum_{i=1}^N\frac{1}{\bar m_i}\chi(\bar m_i \geq m)\Eq{e2.7}
\ee
where $\chi(A)=1$ if $A$ is true and $0$ otherwise. It is convenient to define the probability to get a given $m$ value as
\be
\lis Q(m)=\frac{1}{N}\sum_{i=1}^N\delta(\bar m_i,m)\Eq{e2.8}
\ee
where $\delta(n,m)$ is the Kronecker delta. Once obtained $\lis Q(m)$, we can
get recursively $\bar P(m)$:
\be
\bar P(m+1)=\bar P(m)-\frac{1}{m}\lis Q(m)\quad ,\quad \bar
P(1)=\sum_{m=1}^{\infty}\frac{1}{m}\lis Q(m)=
\frac{1}{N}\sum_{i=1}^N\frac{1}{\bar m_i}\Eq{e2.9}
\ee
and the momenta distribution is then given by:
\be
\s_m=h^m\frac{1}{N}\sum_{i=1}^N\frac{1}{\bar m_i}\sum_{l=1}^{\bar m_i}l^m
\Eq{e2.10}\ee
Finally, the error bars of a probability distribution (for instance $\bar P$) are computed by considering that the probability that in $N$ elements of a sequence there are $n$ in the box $m$ is given by the binomial distribution:
\be
D_m(n,N)=\binom{N}{n} \bar P(m)^n\left(1-\bar P(m) \right)^{N-n}
\Eq{e2.11}\ee
From it we find
\be
\langle n\rangle_m=N\bar P(m)\quad,\quad \langle(n-\langle n\rangle_m)^2\rangle_m= \bar P(m)\left(1-\bar P(m) \right)/N
\Eq{e2.12}\ee
where $\langle .\rangle=\sum_n . D_m(n,N)$. Therefore, the error estimation
for the $\bar P$ probability is given by
\be
\bar P(m) \pm 3 \left(\bar P(m)\left(1-\bar P(m) \right)/N\right)^{1/2}
\Eq{e2.13}\ee
We have done $15$ simulations with $\kappa=22$, $N=10^{12}$ realizations
($10^4$ cycles of size $10^7$) and different values of $W_0$: $10^7$,
$10^8$, $5\times10^8$, $10^9$, $2\times 10^9$, $3\times 10^9$, $4\times
10^9$, $5\times 10^9$, $10^{10}$, $2\times 10^{10}$, $5\times 10^{10}$,
$7\times 10^{10}$, $10^{11}$, $2\times 10^{11}$ and $5\times 10^{11}$. We
also use the Reynold's number $R=W_0^{1/3}$.

\begin{figure}[!h]
\begin{center}
\includegraphics[width=12cm,clip]{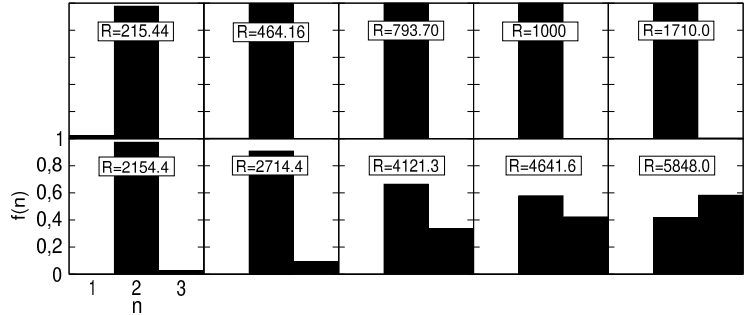}
\end{center}
\kern -2.5cm
\caption{Distribution of events that reach the Kolmogorov scale
  $\kappa^{-n}$ for different values of the Reynold's numbers $R$ and
  $\kappa=22$. The total number of events is $10^{12}$. }
\end{figure}
\begin{figure}[!h]
\begin{center}
\includegraphics[width=10cm,clip]{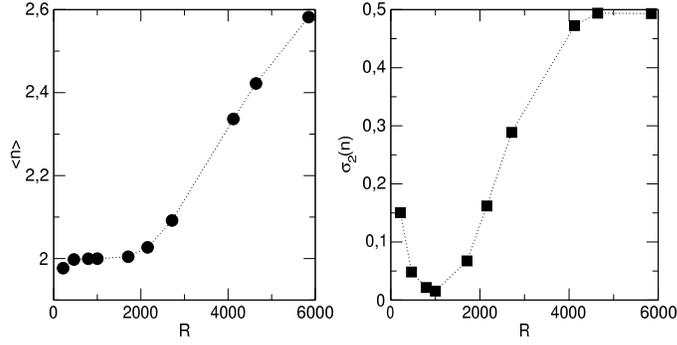}
\end{center}
\kern -2.5cm
\caption{Momenta of the $\bar n$-distribution. 
Left: average value. Right: second momenta}
\end{figure}
\begin{figure}[!h]
\begin{center}
\includegraphics[width=12cm,clip]{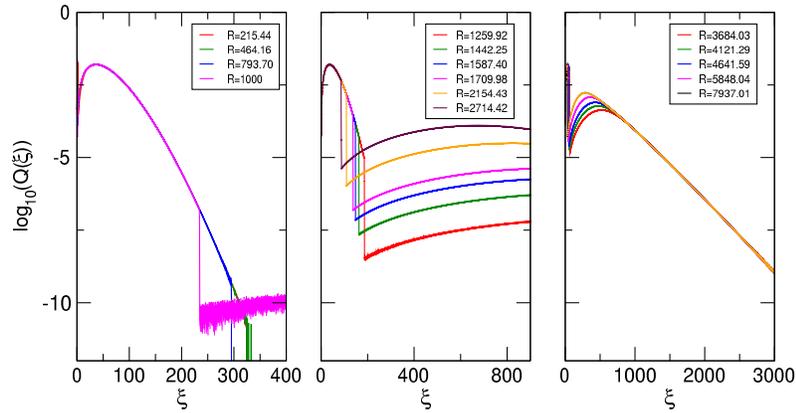}
\end{center}
\kern -1.5cm
\caption{Plot as a function of $\x=m h$
of the logarithm of the probability, $\log_{10} Q(\x)$, with $Q(\x)=\lis
Q(\frac{\x}h)$,
  that the Kolmogorov scale is reached at scale
  $m=\frac{\xi}h$, for different Reynold's numbers and $h=10^{-3}\media{\x}$.} 
\end{figure}
\begin{figure}[!h]
\begin{center}
\kern -0.5cm
\includegraphics[width=13cm,clip]{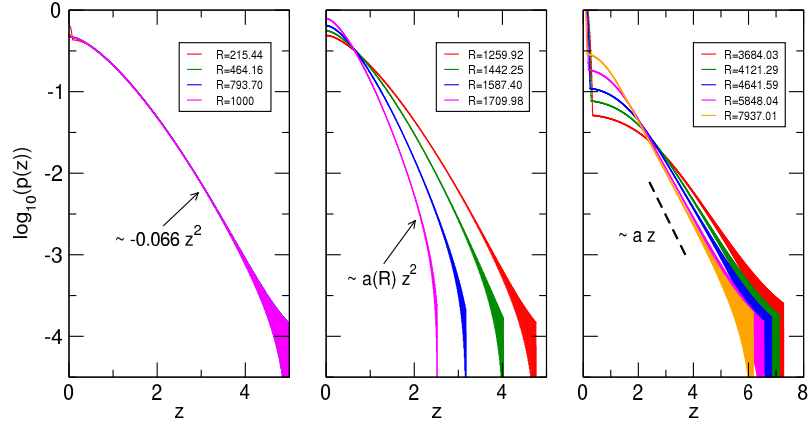}
\end{center}
\kern -2.5cm
\caption{$log_{10}p(z)$ distribution for different Reynold's
  numbers. Central figure: $a(1259.92)=-0.08$, $a(1442.25)=-0.12$,
  $a(1587.40)=-0.24$ and $a(1709.98)=-0.52$. Right figure:
  $a(3684.03)=-0.51$, $a(4121.29)=-0.55$, $a(4621.59)=-0.58$,
  $a(5848.04)=-0.61$ and $a(7937.01)=-0.64$.}
\end{figure}
\begin{figure}[!h]
\kern -0.2cm
\begin{center}
\includegraphics[width=12cm,clip]{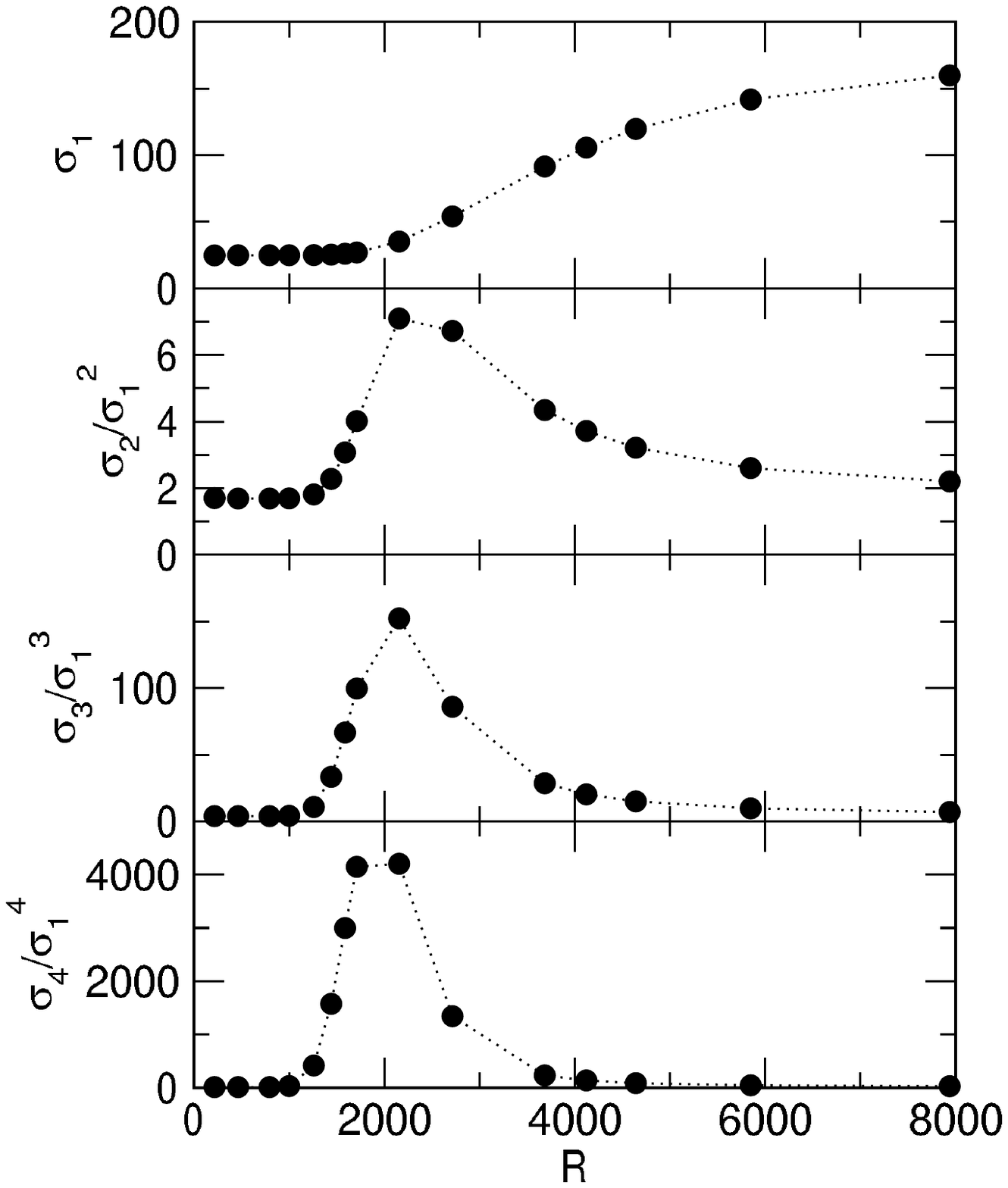}
\end{center}
\kern -1.cm
\caption{Measured momenta of the distribution $P(\xi)$, $\s_n$, vs 
Reynold's number. } 
\end{figure}
The size of $\k$ has been chosen to fit the data for the intermittency
exponents $\z_p$ and it is quite large ($\k=22$, \cite{Ru012}): this has
the consequence that the Kolmogorov scale is reached at a scale $\k^{-n}$
with $n=2,3$ and very seldom for higher scales, at the considered Reynolds
numbers. That can be seen in Figure 1 where we show the obtained
distributions of $\bar n_i$, $i=1,\ldots,N$. In Figure 2 we see the average
value of $\bar n$ and its second momenta. We see that for low Reynold's
numbers the values is almost constant equal to $2$ and from $R\simeq 2000$
it begins to grow. The second momenta shows a minimum for $R\simeq 1000$
where almost all events are in $\bar n_i=2$.

The measured distribution of $Q(\x)$ (see Figure 3) reflects the
superposition of two distributions: the values of $\xi$ associated to the
$\bar n_i=2$ and to $\bar n_i=3$ events. Moreover, for small values of $R$
the overall distribution is dominated by the events $\bar n_i=2$ and for
large Reynold's numbers it is dominated by the $\bar n_i=3$ events. At each
case the form of the distribution is different: for small $R$,
$log_{10}Q(\xi)$ is quadratic in $\xi$ and for large $R$ is linear in
$\xi$.

The behavior of $Q$ defines the behavior of $p(z)$. In Figure 4 we see the
$p(z)$ behavior.  We again see clearly how for low $R$ values the
distribution is non sensitive to the values of $R$ and it is Gaussian. For
intermediate values of $R$ the exponential of a quadratic function is a
good fit for the measured distribution and large enough values of $z$ but
its parameters parameters depend on $R$. Finally for $R$ large of $3000$
the distribution changes and its behavior for large $z$ values seems to be
fitted very well by a linear funcion with a $R$-depending slope.

It is interesting to show the dependence of the momenta of $\bar P$,
$\s_n$, as a function of $R$. In Figure 5 we see their behavior. We can
naturally identify three regions: Region I ($R\in [0,1000]$) where the
momenta are almost constant, Region II ($R\in[1000,4000]$) where the
moments grow with $R$ and Region III ($R\in[4000,\infty]$) where relative
moments tend to some asymptotic value.

\begin{figure}[!h]
\begin{center}
\includegraphics[width=7cm,clip]{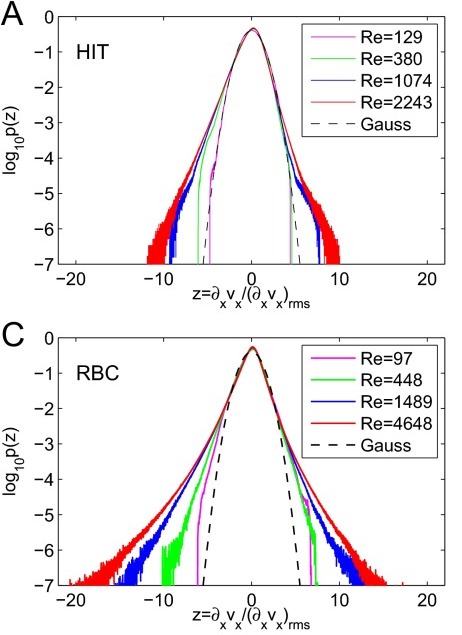}
\end{center}
\caption{Measured $p(z)$ by Schumaher et al. \cite{SSKDYS014} for different Reynold's numbers}
\end{figure}
\begin{figure}[!h]
\begin{center}
\includegraphics[width=10cm,clip]{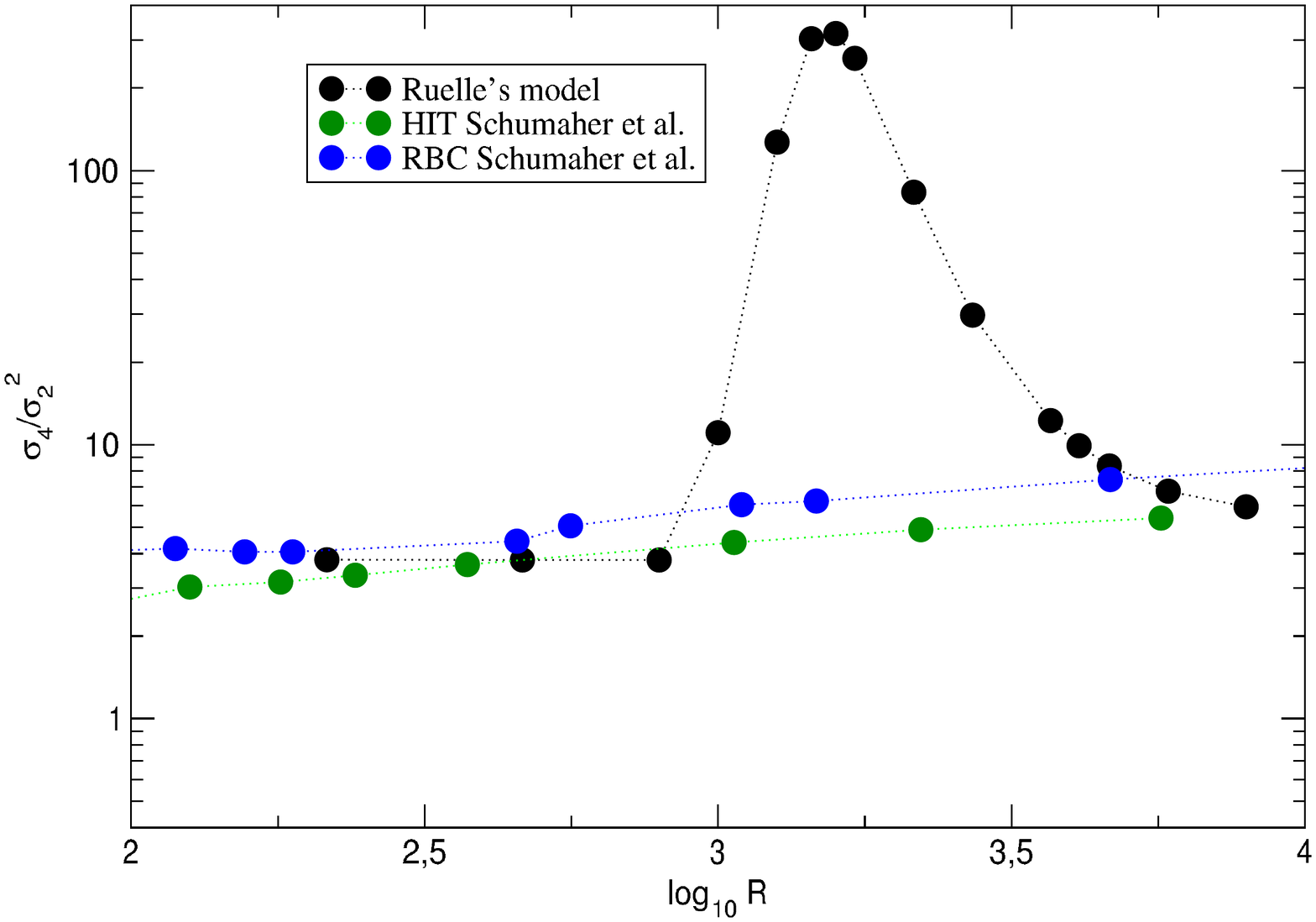}
\end{center}
\kern -1.cm
\caption{Measured flatness ($\s_4/\s_2^2$) compared with the results by Schumaher et al. \cite{SSKDYS014} for different Reynold's numbers}
\end{figure}

Experimental data can be found in \cite{SSKDYS014} and are illustrated by
the two plots in Figure 6 taken from the cited work

\0which give the function $\log_{10}p(z)$. \ie the probability density for
observing a normalized radial gradient $z$ as a function of
$z=\x/\sqrt{\media{\x^2}}$ in the case of homogeneous isotropic turbulence
(HIT) (\ie Navier-Stokes in a cube with periodic boundaries) or in the case
of Raleigh-Benard convection (RBC) (NS+heat transport in a cylinder with
hot bottom and cold top). The results of Fig. 6, for $z>0$ should be
compared with those of Fig.4 at the corresponding Reynolds numbers.  In
both cases we see that the distribution for high Reynold numbers have
linear-like behavior for large $z$-values. In fact for the HIT case and
$R=2243$ we can fit a line with slope $-0.77$ in the interval
$z\in[3.3,5.57]$. Also in the RBC case we can do a linear fit with slope
$-0.42$ ($z\in 5.14,9.69]$) for $R=4648$. The value obtained is similar to
  the ones we computed on Figure 4.

In figure 7, we can compare the measured {\it flatness} in our numerical
experiment with the observed by Schumaher et al. \cite{SSKDYS014}. We see
that the values are similar for small and large Reynold's numbers but there
is a peacked structure for intermediate values due to the relevant
discontinuity when passing from $\bar n_i=2$ to $\bar n_i=3$ events.

All these results shows that the important aspect of the experiments is
quite well captured with the only paramenter $\k$ available for the fits,
\ie a strong deviation from Gaussian behavior and the agreement of the
location of the abscissae of the minima of the tails in the second case at
the maximal $W_0$; this feature fails in the first case (HIT) as the
abscissa is about $30$: is it due to a too small Reynolds number?. This
seems certainly a factor to take into account as the curve appears to
become independent of $R$, hence universal as it should on the basis of the
theory, for $R>4000$.

In conclusion the results are compatible with the OK theory but show
important deviations for large fluctuations because the Gumbel distribution
does not show a Gaussian tail.

All this has a strong conceptual connotation: the basic idea (ie the
proposed hierarchical and scaling distribution of the kinetic energy
dissipation per unit time) is fundamental.

The need to assign a value to the scaling parameter $\k$ is quite
interesting: in the renormalization group studies the actual value of $\k$
is usually not important as long as it is $\k>1$. Here the value of $\k$ is
shown to be relevant (basically it appears explicitly in the end results
and its value $\sim20$ must, in principle, be fixed by comparison with
simulations on fluid turbulence).

\* {\bf Acknowledgements:} The above comments are based on numerical
calculations first done by P. Garrido and confirmed by G. Gallavotti.  This
is the text of our comments (requested by the organizers) to the talk by
D. Ruelle at the CHAOS15 conference, Institut Henri Poincar\'e, Paris, May
26-29, 2015.

\bibliographystyle{plain}

\*
\0{\small Giovanni Gallavotti\hfill\break
INFN-Roma1 and Rutgers University\hfill\break
\tt giovanni.gallavotti@roma1.infn.it}
\hfill\break
{\small Pedro Garrido\hfill\break
Physics Dept., University of Granada\hfill\break
\tt garrido@onsager.ugr.es}

\end{document}